\documentclass[twocolumn,amsmath,amssymb,prx,longbibliography]{revtex4-1}

\usepackage{graphicx}% Include figure files
\usepackage{dcolumn}% Align table columns on decimal point
\usepackage{bm}% bold math
\usepackage{color}
\usepackage{soul}
\usepackage{notes2bib}
\usepackage{ulem }

\begin{document}

%\bibliographystyle{apsrev}

%\preprint{}

\title{All-optical \textit{dc} nanotesla magnetometry using  silicon vacancy fine structure \\ in isotopically purified silicon carbide}

\author{D.~Simin$^{1}$}
\author{V.~A.~Soltamov$^{2}$}
\author{A.~V.~Poshakinskiy$^{2}$}
\author{A.~N.~Anisimov$^{2}$}
\author{R.~A.~Babunts$^{2}$}
\author{D.~O.~Tolmachev$^{2}$}
\author{E.~N.~Mokhov$^{2,5}$}
\author{M.~Trupke$^{3}$}
\author{S.~A.~Tarasenko$^{2}$}
\author{A.~Sperlich$^{1}$}
\author{P.~G.~Baranov$^{2}$}
\author{V.~Dyakonov$^{1,4}$}
\email[E-mail:~]{dyakonov@physik.uni-wuerzburg.de}
\author{G.~V.~Astakhov$^{1}$}
\email[E-mail:~]{astakhov@physik.uni-wuerzburg.de}

\affiliation{$^1$Experimental Physics VI, Julius-Maximilian University of W\"{u}rzburg, 97074 W\"{u}rzburg, Germany \\
$^2$Ioffe Physical-Technical Institute, 194021 St.~Petersburg, Russia\\ 
$^3$Vienna Center for Quantum Science and Technology, Atominstitut, TU Wien, 1020 Wien, Austria\\ 
$^4$Bavarian Center for Applied Energy Research (ZAE Bayern), 97074 W\"{u}rzburg, Germany\\
$^5$St.~Petersburg National Research University of Information Technologies, Mechanics and Optics, 197101 St.~Petersburg, Russia.\\ }

\begin{abstract}
We uncover the fine structure of a silicon vacancy in isotopically purified silicon carbide (4H-$^{28}$SiC) and reveal not yet considered terms in the spin Hamiltonian, originated from the trigonal pyramidal symmetry of this spin-3/2 color center. These terms give rise to additional spin transitions, which would be otherwise forbidden, and lead to a level anticrossing in an external magnetic field. We observe a sharp variation of the photoluminescence intensity in the vicinity of this level anticrossing, which can be used for a purely all-optical sensing of the magnetic field. We achieve $dc$ magnetic field sensitivity better than $100 \, \mathrm{nT / \sqrt{Hz}}$ within  a volume of  $3 \times 10^{-7} \, \mathrm{m m^{3}}$ at room temperature and demonstrate that this contactless method is robust at high temperatures up to at least $500 \, \mathrm{K}$. As our approach does not require application of radiofrequency fields, it is scalable to much larger volumes. For an optimized light-trapping waveguide of $3 \, \mathrm{m m^{3}}$ the projection noise limit is below $100 \, \mathrm{fT / \sqrt{Hz}}$. 
\end{abstract}

\date{\today}

\pacs{76.30.Mi, 71.70.Ej, 76.70.Hb, 61.72.jd}

\maketitle
%---------------------------------------------------------------

\section{Introduction}

The vacancy-related color centers in CMOS-compatible material silicon carbide (SiC) are perspective for chip-scale quantum technologies \cite{Baranov:2007gu,Weber:2010cn,Baranov:2011ib,Riedel:2012jq,Fuchs:2013dz,Castelletto:2013jj,Somogyi:2014jm,Muzha:2014th,Calusine:2014gv} based on ensembles  \cite{Koehl:2011fv,Soltamov:2012ey,Falk:2013jq,Kraus:2013di,Klimov:2013ua,Falk:2014fh,Kraus:2013vf,Yang:2014kqa,Zwier:2015go,Falk:2015iz,Carter:2015vc,Simin:2015dn,Lee:2015ve} as well as on single centers \cite{Castelletto:2013el,Castelletto:2014eu,Christle:2014ti,Widmann:2014ve,Fuchs:2015ii,Lohrmann:2015hd}. Similar to the spin $S = 1$ nitrogen-vacancy (NV) defect in diamond -- which has become a standard solid-state system for the application of  quantum sensing under ambient conditions  \cite{Maze:2008cs,Balasubramanian:2008cz,Wolf:2014us} -- the silicon vacancy ($\mathrm{V_{Si}}$) in SiC possesses selectively addressable spin states through optically detected magnetic resonance (ODMR) \cite{Riedel:2012jq}. Unlike the spin-1 defects, the higher half-integer spin $S = 3/2$ of $\mathrm{V_{Si}}$ \cite{Mizuochi:2002kl,Kraus:2013di} provides additional degree of freedom \cite{Lanyon:2008gvb} and functionality \cite{Simin:2015dn}, but it is usually unutilized.  A major obstacle is that the structure of high-spin centers being far more complicated is not known yet. Meanwhile, it is the level fine structure that is the key to understand spin dynamics and relaxation processes, which sets up limits for the performance of potential devices.

Here, we reveal the fine structure of the $\mathrm{V_{Si}}$ ground and excited states (GS and ES, respectively) in external magnetic fields.  We show that the $C_{3v}$ point group  of the $\mathrm{V_{Si}}$ defect gives rise to additional terms in the spin Hamiltonian, which have not been considered so far. Particularly, the trigonal pyramidal symmetry of the $\mathrm{V_{Si}}$ defect enables spin transitions with a change of the spin projection $\Delta m_S = \pm 2$. As compared to the commonly studied spin transitions with $\Delta m_S = \pm 1$, they are induced by counter circularly polarized radiation and their energies shift with the double slope in a magnetic field.  Moreover, we observe two GS level anticrossings (LAC) between the $m_S = - 3/2$ and both $m_S = -1/2$ (GSLAC-1) and $m_S = +1/2$ (GSLAC-2) spin sublevels. The GSLAC-2 can fundamentally occur for color centers with the spin $S \geq 3/2$ only. We develop a theory of the $\mathrm{V_{Si}}$ fine structure, which precisely takes into account the real atomic arrangement of the vacancy and quantitatively describes the experimental findings. The photoluminescence (PL) intensity demonstrates resonance-like behavior in the vicinity of LACs, and the sharpest resonance is detected for GSLAC-2, determined by the parameters related to the trigonal pyramidal symmetry of the $\mathrm{V_{Si}}$ center. In the following, we show that this optical phenomenon can be used  to measure $dc$ magnetic fields without a need to apply radiofrequency (RF) fields and we demonstrate nanotesla resolution within sub-1000 $\mathrm{\mu m^{3}}$. 
The effect is robust up to at least $500 \, \mathrm{K}$, suggesting a simple, contactless method to monitor weak magnetic fields in a broad temperature range. 

Our approach is easily-scalable, and for a probe volume on the order of $1 \, \mathrm{m m^{3}}$ with improved pump/collection efficiency, we expect magnetic field sensitivity to be about hundred femtotesla per square root of Hertz. While coming close to the sensitivity of other benchmark chip-scale magnetic field sensors \cite{Clevenson:2015cv,Shah:2007kb}, this technique relies neither on RF fields, as for the NV defects in diamond \cite{Clevenson:2015cv}, nor on vapour heating, as for the microfabricated rubidium cells \cite{Shah:2007kb}. Furthermore, the proposed method is not restricted to magnetic sensing and can potentially be extended for radiofrequency-free sensing of other physical quantities, particularly temperature and axial stress.

\section{Experiment}

\begin{figure*}[t]
\includegraphics[width=.59\textwidth]{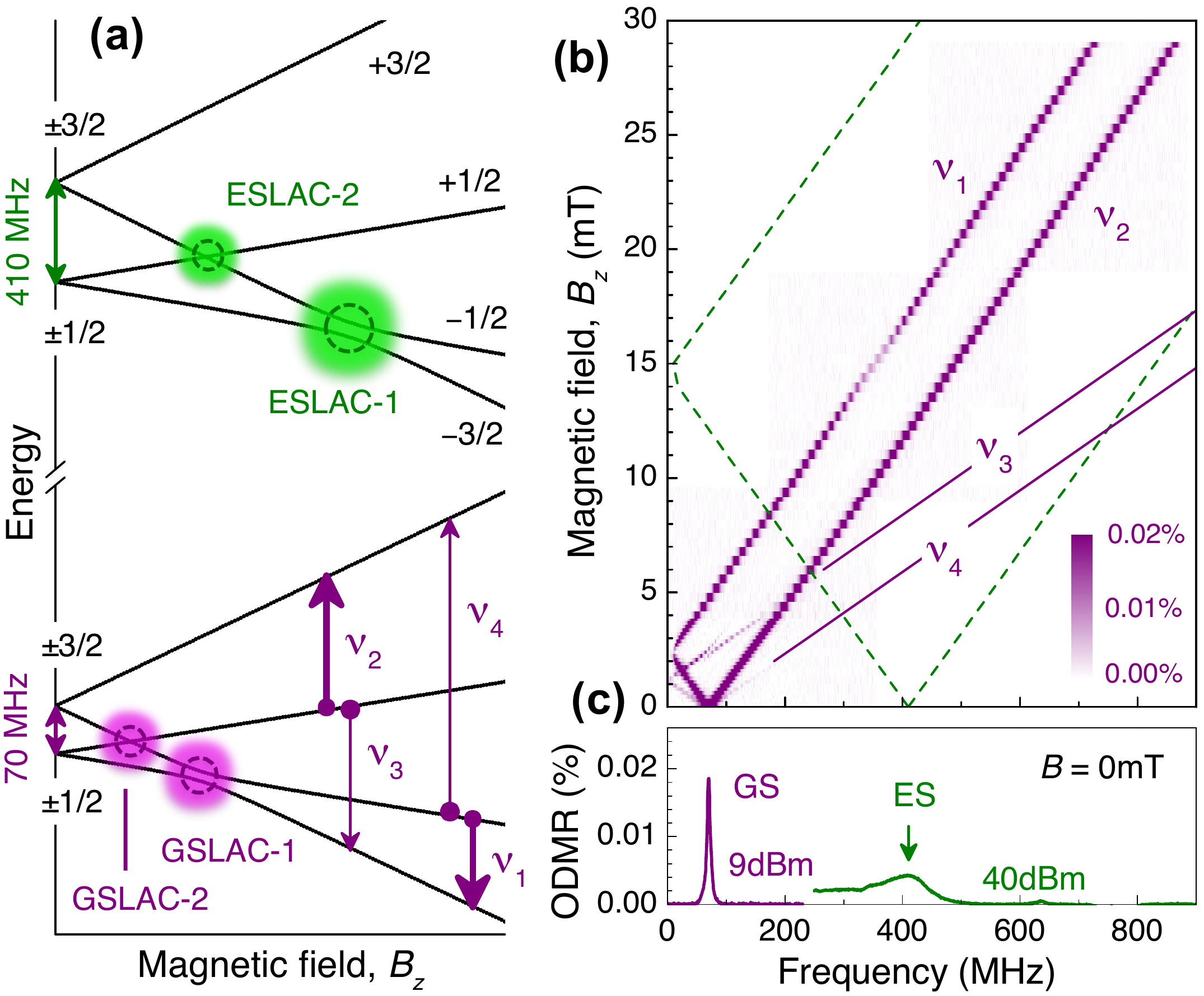}
\caption{ (a) The GS ($2D = 70 \, \mathrm{MHz}$) and ES ($2D' = 410 \, \mathrm{MHz}$) spin sublevels of the $\mathrm{V_{Si}}$ point defect in external magnetic field $ B_z \| c$, assuming a weak perpendicular component $B_{\bot} \ll B_z$. The vertical arrows indicate RF driven spin transitions, their thicknesses mirroring the contrasts of the corresponding ODMR lines. (b) Magnetic field versus frequency evolution of the $\mathrm{V_{Si}}$ ODMR signal recorded at room temperature and at low RF power. The solid and dashed lines are the positions of the ODMR peaks calculated for the $\Delta m_S = \pm 2$ transitions in the GS and  $\Delta m_S = \pm 1$ transitions in the ES, respectively. (c) Low-RF-power ($9 \, \mathrm{dBm}$) and high-RF-power ($40 \, \mathrm{dBm}$) ODMR spectra in zero magnetic field. } \label{fig1}
\end{figure*}

In natural SiC, the ODMR spectra of the $\mathrm{V_{Si}}$ defects are affected by the hyperfine interaction with the $^{29}$Si-isotope nuclear spin $I = 1/2$ \cite{Sorman:2000ij,Kraus:2013di}. In order to elude this interaction, we use isotopically purified SiC with above 99.0\% of $^{28}$Si nuclei with $I = 0$. To obtain such a crystal, we first synthesize polycrystalline SiC with the use of silicon and carbon powders, the former being enriched with the $^{28}$Si isotope. The polycrystalline substance is then used as a source for the growth of 4H-$^{28}$SiC crystals by the sublimation method in a tantalum container  \cite{Karpov:2000jl}. The growth is performed in vacuum on 4H-SiC substrates at a temperature of 2000$^\circ$C. The growth rate is approximately $0.25 \, \mathrm{mm / hour}$. Afterwards, we polish out the substrate, obtaining the sample with a thickness of about $500 \, \mathrm{\mu m}$. In order to introduce the silicon vacancies, the sample is irradiated with neutrons in a nuclear reactor with a fluence of $1 \times 10^{16} \, \mathrm{cm^{-2}}$, resulting in a nominal $\mathrm{V_{Si}}$ density of $2 \times 10^{14} \, \mathrm{cm^{-3}}$ \cite{Fuchs:2015ii}.

To optically address the $\mathrm{V_{Si}}$ spin states, we use a 785-$\mathrm{nm}$ laser diode. The optical excitation followed by the spin-dependent recombination leads to a preferential population of the $m_S = \pm 1/ 2$ sublevels along the crystal symmetry $c$-axis \cite{Baranov:2007gu}. The PL from $\mathrm{V_{Si}}$ occurs in the near infrared spectral range \cite{Hain:2014tl}, and it is selected and detected using a Si photodiode and a 900-nm long-pass filter. The PL intensity is spin dependent: in case of the $\mathrm{V_{Si}}$ center studied here -- so called V2 center -- it is higher when the system is in the $m_S = \pm 3/ 2$ states and lower when the system is in the $m_S = \pm 1/2$ states \cite{Sorman:2000ij, Baranov:2011ib,Kraus:2013vf}. The laser beam is focused onto the sample using a 20$\times$ optical objective ($\mathrm{N.A. }= 0.3$), optimized for the near infrared light, and the PL is collected through the same objective. The nominal excitation volume is $330 \, \mathrm{\mu m^{3}}$. To additionally manipulate the $\mathrm{V_{Si}}$ spin states, we apply RF field, provided by a signal generator. RF is then amplified, guided to a  500-$\mathrm{\mu m}$-thick stripline and terminated with a 50-$\Omega$  impedance. A static magnetic field can be applied in an arbitrary direction, using a 3D coil arrangement in combination with a permanent magnet. The field direction and strength are calibrated using a 3D Hall sensor.  

\begin{figure*}[t]
\includegraphics[width=.61\textwidth]{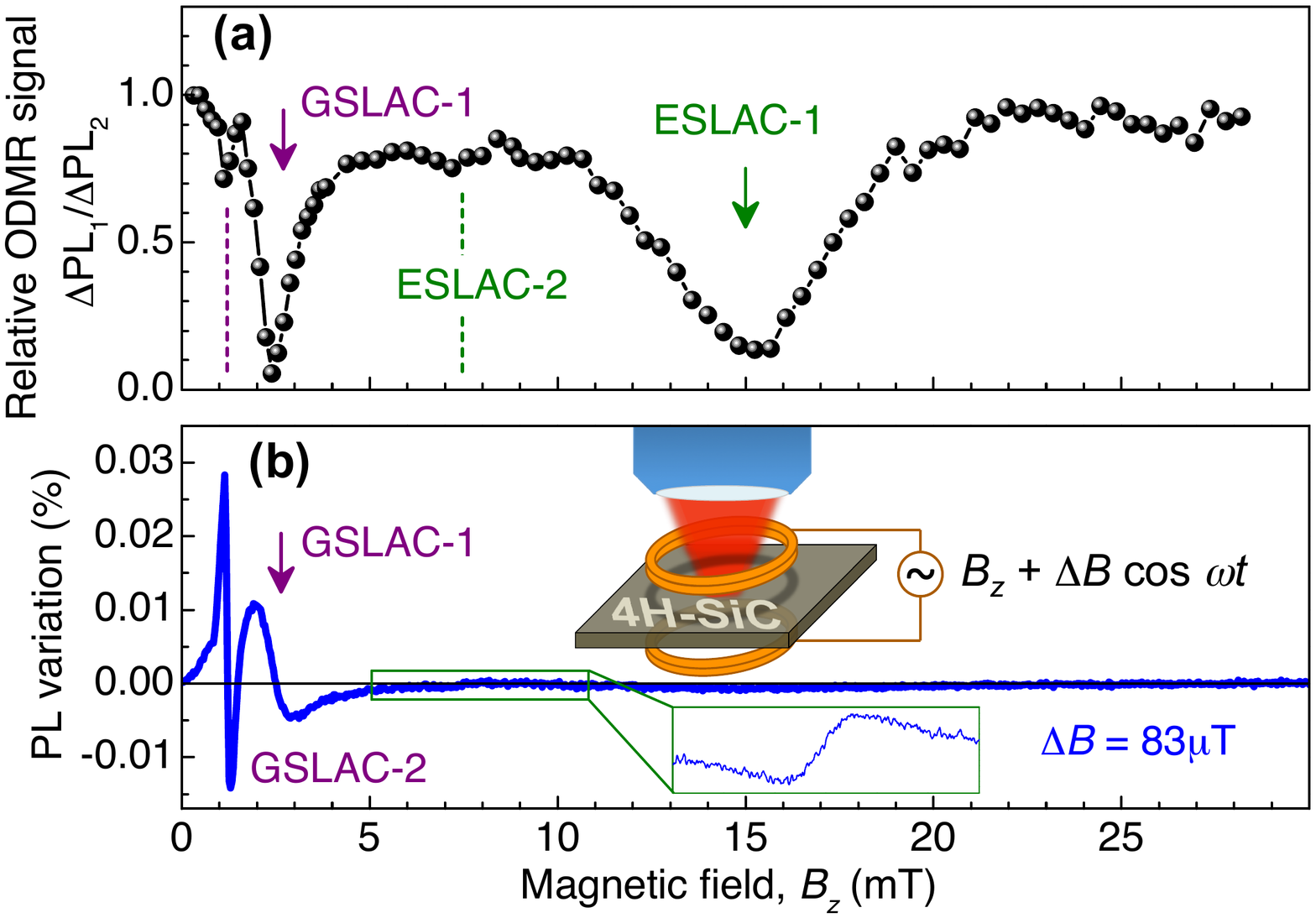}
\caption{ (a) Relative strength of the $\nu_1$ and $\nu_2$ ODMR transitions ($\mathrm{\Delta PL_1 / \Delta PL_2}$) as a function of the magnetic field $\mathbf{B}$ applied parallel to the $c$-axis of 4H-SiC. The arrows indicate the positions of GSLAC-1 ($2.5 \, \mathrm{mT}$) and ESLAC-1 ($15 \, \mathrm{mT}$). The vertical dashed lines correspond to the expected positions of GSLAC-2 (lower field) and ESLAC-2 (higher field). (b) Lock-in detection of the PL variation $\mathrm{\Delta PL / PL}$ as a function of the $dc$ magnetic field $B_z$, where $\mathrm{\Delta PL}$ is caused by the application of an additional weak oscillating magnetic field $\Delta B$, i.e., $B_z + \Delta B  \cos \omega t$ with $\Delta B = 83 \, \mathrm{\mu T}$ and $\omega / 2 \pi = 5 \, \mathrm{kHz}$. The sharp resonance at $1.25 \, \mathrm{mT}$  corresponds to GSLAC-2. Resonant RF field is not applied. Upper inset: A scheme of the experiment.  Lower inset: A detailed measurement in the magnetic field range corresponding to ESLAC-2. } \label{fig2}
\end{figure*}

In the absence of an external magnetic field, the $\mathrm{V_{Si}}$ GS is split in two Kramers degenerate spin sublevels $m_S = \pm 3/2$ and $m_S = \pm 1/2$ with the zero-field splitting $2D = 70 \, \mathrm{MHz}$ \cite{Sorman:2000ij,Kraus:2013vf}. When an external magnetic field $\mathbf{B}$ is applied parallel to the $c$-axis, the spin states are further split and the splitting is linear with $B_z$ ($z \| c$), as schematically shown in Fig.~\ref{fig1}(a). A resonance RF induces magnetic dipole transitions between the spin-split sublevels $(\pm1/2  \rightarrow \pm3/2)$, resulting in a change of the PL intensity ($\mathrm{\Delta PL}$). The room-temperature evolution of the ODMR spectrum (i.e., the RF-dependent ODMR contrast $\mathrm{\Delta PL / PL}$) with the external magnetic field $B_z$ is presented in Fig.~\ref{fig1}(b). 

We first discuss the case of $B_z = 0$ [Fig.~\ref{fig1}(c)]. At a low RF power of $9 \, \mathrm{dBm}$, we detect a single ODMR line at a frequency $\nu_0 = 70 \, \mathrm{MHz}$, which is equal to $2D$ in the GS. At much higher RF power ($40 \, \mathrm{dBm}$), we detect another ODMR line at a frequency of $410 \, \mathrm{MHz}$. Below, we establish that it corresponds to the zero-field splitting $2D'$ in the ES [Fig.~\ref{fig1}(a)]. 
We would like to mention that similar resonance was observed previously \cite{Kraus:2013vf}  and ascribed to the Frenkel pair \cite{vonBardeleben:2000jg}. 

Upon application of a magnetic field $B_z$, one of the RF driven transitions is $(-1/2 \rightarrow -3/2)$ with $\Delta m_S = - 1$, and the corresponding ODMR line $\nu_1 = | \nu_0 - g_{\parallel}   \mu_B B_z / h|$ shifts linearly with $B_z$. Another RF-driven transition is $(+1/2  \rightarrow +3/2)$ with $\Delta m_S = + 1$,  and the corresponding ODMR line $\nu_2 =  \nu_0 + g_{\parallel}  \mu_B B_z / h$ shifts linearly towards higher frequencies. These two transitions are indicated by the thick arrows in Fig.~\ref{fig1}(a) and the corresponding ODMR lines are clearly seen in Fig.~\ref{fig1}(b), being in agreement with the previous results \cite{Kraus:2013vf,Simin:2015dn}. Remarkably, at a magnetic field $B_{G1} = h \nu_0 / g_\parallel   \mu_B = 2.5 \, \mathrm{mT}$ the frequency of the $\nu_1$ ODMR line tends to zero [Fig.~\ref{fig1}(b)], which is due to the GSLAC-1 [Fig.~\ref{fig1}(a)]. 

We analyze the relative contrast of the $\nu_1$ and $\nu_2$ ODMR lines as a function of $B_z$ and observe two pronounced dips [Fig.~\ref{fig2}(a)]. One of them is at $B_{G1} = 2.5 \, \mathrm{mT}$ (i.e., exactly at GSLAC-1) and the other one is at $B_{E1} = 15 \, \mathrm{mT}$. The dashed lines in Fig.~\ref{fig1}(b) represent the calculated evolution of the  ODMR spectrum associated with the $2D' = 410 \, \mathrm{MHz}$ resonance assuming the effective $g$-factor  $g_{\parallel} \approx 2.0$. As expected, the ESLAC-1 occurs at $B_{E1}$. We hence can reconstruct the ES spin structure, as shown in Fig.~\ref{fig1}(a). It agrees with the conclusion drawn from another recent experiment \cite{Carter:2015vc}. The observation of the dip at $15 \, \mathrm{mT}$ in the $\nu_1$ rather than in the $\nu_2$ ODMR signal unambiguously determines the order of the spin sublevels in the ES, i.e., the $m_S = \pm 3/2$ state has higher energy than the $m_S = \pm 1/2$ state ($D' > 0$). 

The appearance of dips in the ODMR signal of Fig.~\ref{fig2}(a) is explained by modification of the optical pumping cycle in the vicinity of LACs either in the GS or ES, which, in turn, results in a change of the PL intensity, as previously reported for some other systems and techniques \cite{vanOort:1991ik,Martin:2000ft,LAC2001,Epstein:2005fi, Rogers:2009hn,Tetienne:2012fv}. 
This suggests that LACs can be detected even without application of RF, simply by monitoring the PL intensity as a function of $B_z$. A scheme of this experiment is presented in the inset of Fig.~\ref{fig2}(b). In order to increase the sensitivity,  we modulate the $dc$ magnetic field $B_z$ by additionally applying a small oscillating field $ \Delta B  \cos \omega t$ from the Helmholtz coils. The correspondingly oscillating PL signal detected by a photodiode is locked-in, mirroring the first derivative of the PL on $B_z$. The experimental curve, recorded at a modulation frequency $\omega / 2 \pi = 5 \, \mathrm{kHz}$ with a modulation depth $\Delta B = 83 \, \mathrm{\mu T}$,  is presented in Fig.~\ref{fig2}(b). Surprisingly, in addition to the GSLAC-1 we detect a pronounced resonance-like behaviour around $B_{G2} = 1.25 \, \mathrm{mT}$. %associated with the GSLAC-2. 

\begin{figure}[t]
\includegraphics[width=.48\textwidth]{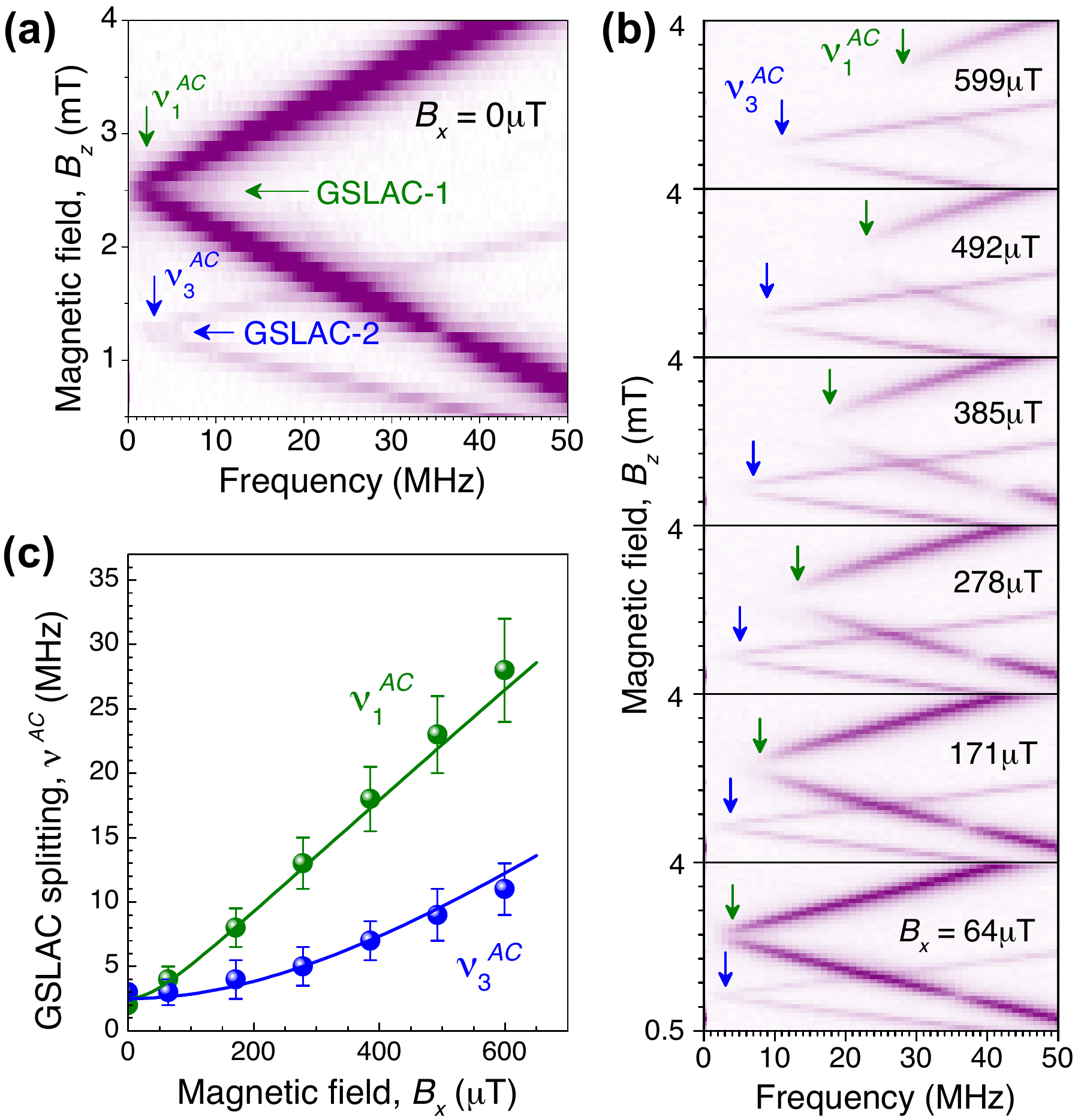}
\caption{ (a) Evolution of the $\nu_{1}$ and $\nu_{3}$ ODMR lines in the vicinity of GSLACs as a function of the magnetic field $B_z$. The perpendicular components of the geomagnetic field are  compensated, $B_x = B_y =0 \, \mathrm{\mu T}$.  (b) Same as (a), but measured for non-zero $B_x$. The vertical arrows indicate the positions of the turning points $\nu_{1}^{AC}$ and $\nu_{3}^{AC}$, which are a direct measure of the level splittings at GSLAC-1 and GSLAC-2, respectively. (c) The GSLAC splittings as a function of $B_x$. The solid lines are calculations as explained in the text. } \label{fig3}
\end{figure}

In order to understand the origin of two GSLACs, we measure the evolution of the ODMR spectrum as a function of $B_z$ with higher precision, in particular, with compensated transverse components of the geomagnetic field, $B_x = B_y =0 \, \mathrm{\mu T}$. The zoom-in of the spectral evolution in the vicinity of GSLAC-1 and GSLAC-2 is shown in Fig.~\ref{fig3}(a). Beside the $\nu_1$ line, the ODMR spectrum contains an additional ($\nu_3$) line, corresponding to the spin transition between the $m_S = + 1/2$ and $m_S = - 3/2$ GSs [sketched by a thin arrow in Fig.~\ref{fig1}(a)]. The turning points of the $\nu_1$ and $\nu_3$ lines correspond to GSLAC-1 and GSLAC-2, respectively. The corresponding spectral shifts of the turning points $\nu_1^{AC}$ and $\nu_3^{AC}$, which are direct measures of the splitting values, are clearly detectable in Fig.~\ref{fig3}(b) for various perpendicular component $B_\perp$ of the magnetic field. The level splittings at both GSLAC-1 and GSLAC-2 grow linearly with $B_\perp$,  but with different slopes [Fig.~\ref{fig3}(c)]. We emphasize that the amplitude of the $\nu_3$ line would rapidly tend to zero and the GSLAC-2 would disappear in an uniaxial model of the defect for magnetic fields being only slightly misaligned from the $c$-axis and hence should be vanishingly small in the experiment, assuming this model. We present below the exact calculation of the relative spin transition rates. Contrarily, the  $\nu_{3}$ ODMR line is clearly detectable in Fig.~\ref{fig3}(a) with the relative strength of the ODMR transition $\mathrm{\Delta PL_3 / \Delta PL_1} =   0.12 \pm 0.02$ and the most pronounced feature in Fig.~\ref{fig2}(b) relates to the GSLAC-2. 

While the magnetic disorder caused by hyperfine interaction with the residual $^{29}$Si nuclei or inaccurate orientation
of the external magnetic field along the $c$-axis can in principle give rise to GSLAC-2 and the $\nu_3$ line, we conclude that
they are not responsible for the above effects. Calculations (see Appendix~\ref{Unimodel}) yield that, for the average nuclear field seen by the $\mathrm{V_{Si}}$ centers of about $( h / 2 \mu_B ) \times 1$\,MHz  \cite{Mizuochi:2002kl, Soltamov:2015prl} or magnetic field misalignment of 
$1^{\circ}$, the contrast ratio between the $\nu_3$ and $\nu_1$ ODMR lines is estimated to be $10^{-3}$, which is by two orders of magnitude lower than that observed in our experiments. Moreover, we observe that the amplitude of the $\nu_3$ line is the same in the natural (presented below) and isotopically purified SiC samples, while the abundance of the spin-carrying $^{29}$Si nuclei differs by a factor of five.
Considering these results, a whole new approach to the spin structure of the $\mathrm{V_{Si}}$ defect is needed and will be thoroughly assembled below.

\begin{table}[h] %[tdph]
\caption{The $g$-factors in Hamiltonian~\eqref{H} which, together with the zero-field splitting $2D = 70 \, \mathrm{MHz}$ and $g_{\parallel} \approx g_{\perp} \approx 2.0$, describe the GS fine structure of the silicon vacancy (V2 center in 4H-SiC) in a magnetic field. }
\begin{center}
\begin{tabular}{|c|c|c|c|}
$g_{2\parallel}$ & $g_{2\perp}$ & $\,\, g_{3\perp}+ g_{3\parallel}/2 \,\,$ & $\,\, g_{3\perp}- g_{3\parallel}/2 \,\, $ \\ \hline 
$\, 0.0 \pm 0.1 \,$ & $\, 0.0 \pm 0.1 \,$ & $\,  0.5 \pm 0.2 \,$ & $\, -0.1 \pm 0.4 \,$   \\ 
\end{tabular}
\end{center}
\label{AllParam}
\end{table}

\section{Silicon vacancy fine structure}

\begin{figure}[t]
\includegraphics[width=.47 \textwidth]{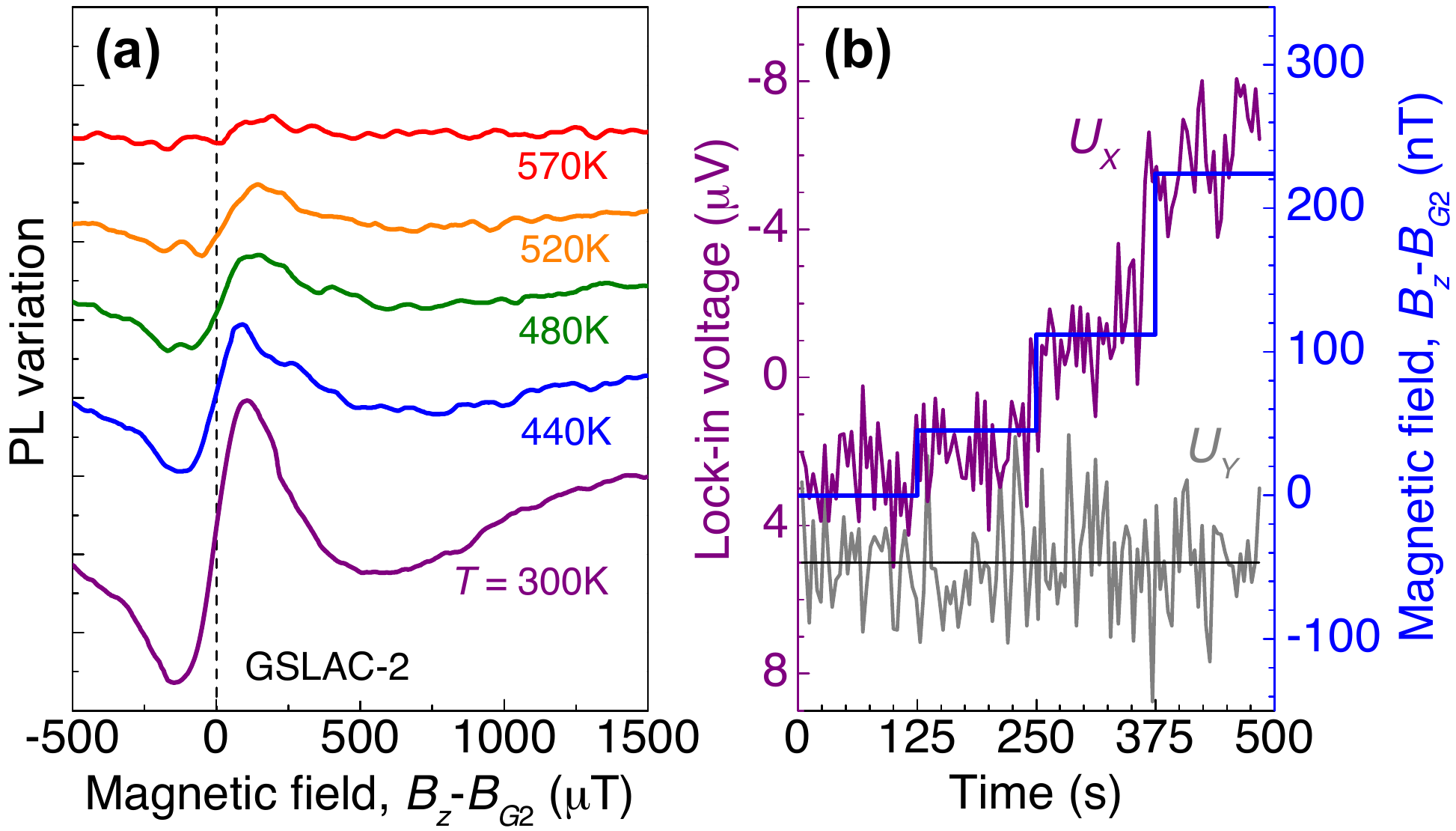}
\caption{ (a) Lock-in detection of the PL variation in the vicinity of GSLAC-2 ($B_{G2} = 1.25 \, \mathrm{mT} $) under application of a weak oscillating magnetic field, recorded at different temperatures. (b)  The in-phase $U_X$ and quadrature $U_Y$ components of the lock-in photovoltage (left axis) for different magnetic fields (right axis), increased in sub-$\mathrm{\mu T}$ steps every $125 \, \mathrm{s}$. The horizontal line is the $U_Y$ mean value. The maximum field sensitivity is obtained for  $\Delta B = 200 \, \mathrm{\mu T}$ and $\omega / 2 \pi = 511 \, \mathrm{Hz}$. Temperature in (b) is $T = 300 \, \mathrm{K}$.} \label{fig4}
\end{figure}

Our findings can be explained in the framework of the spin Hamiltonian, which precisely takes into account the real microscopic $C_{3v}$ group symmetry of the defect \cite{Mizuochi:2003di}. The effective Hamiltonian to the first order in the magnetic field can be presented as a sum of three contributions 
\begin{equation}\label{sum}
\mathcal{H}=\mathcal{H}_0+\mathcal{H}_{1\parallel}+\mathcal{H}_{1\perp} \,,
\end{equation}
where $\mathcal{H}_0$ is the Hamiltonian in zero magnetic field, $\mathcal{H}_{1\parallel} \propto B_z$ and $\mathcal{H}_{1\perp} \propto \mathbf{ B}_\perp = (B_x,B_y)$ are the magnetic-field-induced terms. The Hamiltonians $\mathcal{H}_0$, $\mathcal{H}_{1\parallel}$, and $\mathcal{H}_{1\perp}$ can be constructed applying the theory of group representations~\cite{BP_book}, see Appendix~\ref{Hamiltonian_construction} for details. In the $C_{3v}$ group,
the magnetic field component $B_z$ and the spin operator $S_z$ transform under the irreducible representation $A_2$,
the pairs of the in-plane components $(S_x, S_y)$ and $(B_x, B_y)$ transform under the representation $E$, and the Hamiltonian must be invariant (representation $A_1$). Using the multiplication table for the representations, one can construct all possible invariant combinations of the magnetic field components and the first, second and third powers of the spin operator components.
The forth and higher powers of the spin-$3/2$ operator can be reduced to the operators of lower powers. Finally, taking into account that the Hamiltonian must be invariant with respect to the time reversal and, therefore, $\mathcal{H}_0$ is even in $\mathbf{ S}$ while $\mathcal{H}_{1\parallel}$ and $\mathcal{H}_{1\perp}$ are odd in $\mathbf{ S}$, we obtain
\begin{align}\label{H}
 &\mathcal H_0 = D \left( S_z^2 - \frac{5}{4} \right) ,\nonumber\\\nonumber
  &\mathcal H_{1\parallel} =  \left[ g_{\parallel} S_z +  g_{2\parallel} S_z \left(S_z^2 - \frac{5}{4} \right) + g_{3\parallel}\frac{S_+^3 - S_-^3 }{4{\rm i}} \right] \mu_B  B_z , \\
   &\mathcal H_{1\perp} =   g_{\perp}\mu_B \mathbf{S}_\perp \cdot \mathbf{ B}_\perp +  2g_{2\perp}\mu_B \left\{ \mathbf{ S}_\perp \cdot \mathbf{ B}_\perp, S_z^2-\frac{3}{4} \right\}
 \nonumber \\ &+ g_{3\perp} \mu_B \frac{\{S_+^2,S_z \} B_+ - \{S_-^2,S_z\} B_- }{2{\rm i}}  \,.
\end{align}
Here, $S_x,S_y,S_z$ are the spin-$3/2$ operators, $\bm S_\perp = (S_x,S_y)$, $S_\pm = S_x \pm {\rm i} S_y$, $B_\pm = B_x \pm {\rm i} B_y$, $\{A,B\}=(AB+BA)/2$ is the symmetrized product, $z$ is parallel to $c$-axis, $x$ and $y$ are the perpendicular axes with $y$ lying in a mirror reflection plane, and $\mu_B$ is the Bohr magneton.
Six $g$-factors introduced in Eq.~\eqref{H} are linearly-independent in a structure of the $C_{3v}$ point group.
They can be determined from experimental data, as we do below, or obtained from ab-initio calculations, which is out of scope of the present manuscript.
The  difference $g_{\parallel}-g_{\perp}$ as well as the non-zero values of $D$, $g_{2\parallel}$ and $g_{2\perp}$ are due to non-equivalence of the $z$ axis and the perpendicular axes. The $g$-factors $g_{3\parallel}$ and $g_{3\perp}$ emerge due to the trigonal pyramidal symmetry of the defect.
Hamiltonian~\eqref{H} can also be presented in the equivalent matrix form (Appendix~\ref{Matrix_form}).
%\cite{Averkiev:1981,Durnev:2013}

The parameters of Hamiltonian~\eqref{H} can be determined from the experimental data.
First from the ODMR lines  in the parallel $B_z$ [Fig.~\ref{fig1}(b)] and perpendicular $B_x$ (Appendix~\ref{Extracting_parameters}) magnetic fields, we confirm that to the second digit accuracy $g_{\parallel}, g_{\perp}  \approx 2.0$, in agreement with earlier studies \cite{Sorman:2000ij}, and $|g_{2\parallel}|, |g_{2\perp}| \ll 1$. Then, using a procedure that is independent of the values of $g_{\parallel}$ and $g_{\perp}$, we estimate the ratios $g_{2\parallel}/g_{\parallel}$ and $g_{2\perp}/g_{\perp}$  using Eqs.~\eqref{Bration} and \eqref{Slope} (Appendix~\ref{Extracting_parameters}). 

Furthermore, the Hamiltonian~\eqref{H} describes the opening of spectral gaps due to the perpendicular field component $B_\perp$ at GSLAC-1 and  GSLAC-2.
The corresponding splittings $\Lambda_1$ and $\Lambda_2$ scale linearly with the perpendicular field for $\mu_B B_\perp \ll D$.
Up to linear terms in $g_{2\parallel},g_{2\perp}$ and quadratic terms in $g_{3\parallel},g_{3\perp}$ the splittings in small fields are given by 
\begin{equation}
 \begin{split}
 & \Lambda_1 \approx \sqrt{3} \left(  1 +  \frac{g_{2\perp}}{g_{\perp} } - \frac{g_{3\parallel}g_{3\perp} }{ g_{\parallel} g_{\perp}} -   \frac{g^2_{3\parallel}}{8g^2_{\parallel}}   \right)  g_{\perp} \mu_B B_\perp  \,, \\
 & \Lambda_2 \approx \sqrt{3} \left(   \frac{g_{3\perp}}{g_{\perp}} + \frac{ g_{3\parallel}}{2g_{\parallel}} \right) g_{\perp}  \mu_B B_\perp  \,.
 \end{split}
 \label{SplittingLAC}
\end{equation}
The GSLAC-2 emerges due to the trigonal asymmetry of the silicon vacancy, and the corresponding energy splitting is expected to be smaller than that in the GSLAC-1, $ \Lambda_2 <  \Lambda_1$. Exactly such a behavior is observed in  the experiment of Fig.~\ref{fig3}. We fit the positions of the turning points as $h \nu_{1,3}^{AC} = [\Lambda_{1,2}^2(B_x) + \Lambda_{0}^2]^{1/2}$, where $\Lambda_{0} / h = 2.5 \, \mathrm{MHz}$ accounts for finite ODMR linewidth and inhomogeneity. From the best fit [the solid lines in Fig.~\ref{fig3}(c)], we first obtain $g_{3\perp}+g_{3\parallel}/2 =0.5 \pm 0.2$ using the data for $\Lambda_{2}$ and then estimate the value $g_{3\perp}-g_{3\parallel}/2=-0.1\pm0.4$  using the data for $\Lambda_{1}$. All the $g$-factors of the Hamiltonian~\eqref{H} are summarized in table~\ref{AllParam}. It is instructive to compare the results with a high-symmetry defect of the $T_d$ point group, where one  expects the relation $g_{3\parallel}/g_{3\perp}=2/3$ (Appendix~\ref{Matrix_form}). 

We are now in the position to explain the appearance of the $\nu_3$ and $\nu_4$ ODMR lines in Figs.~\ref{fig1}(b), \ref{fig3}(a) and \ref{fig3}(b), even when $B_\perp = 0$. It follows from the Hamiltonian~\eqref{H} that the matrix elements of the allowed magnetic dipole transitions have the form
\begin{align}
  & M_{\text{$\mp 3/2$, $\mp 1/2$}} = \frac{\sqrt{3}}{2} \left( 1 + \frac{g_{2\perp}}{g_{\perp}} \right) g_{\perp} \mu_B B_{1,\sigma^{\mp}}  \,,  \label{M1} \\
   & M_{\text{$\mp 3/2$, $\pm 1/2$}} = - {\rm i} \frac{\sqrt{3}}{2} \left(  \frac{g_{3\perp}}{g_{\perp}} + \frac{g_{3\parallel}}{2g_{\parallel}} \right)  g_{\perp}  \mu_B B_{1,\sigma^{\pm}}  \,, \label{M2}
\end{align}
where $\mathbf{ B}_1$ is the RF magnetic field and $B_{1,\sigma^{\pm}}  = B_{1,x} \mp {\rm i} B_{1,y}$. The transitions $(+1/2  \rightarrow -3/2)$ and $(-1/2  \rightarrow +3/2)$, responsible for the $\nu_3$ and $\nu_4$ ODMR lines, respectively, occur due to the trigonal pyramidal symmetry of the spin-3/2 defect and are induced by the $\sigma^{+}$ and  $\sigma^{-}$ circularly polarized RF radiation. There are two microscopic contributions to these transitions: (i) coupling of the $m_S = +3/2$ and $m_S = -3/2$ states by the longitudinal static field $B_z$ (parameter $g_{3\parallel}$)~\cite{Averkiev:1981,Durnev:2013gi} followed by the RF driven transitions with $\Delta m = \pm 1$ and (ii) direct coupling of the $m_S = +3/2$ and $m_S=-1/2$ as well as $m_S = -3/2$ and $m_S = +1/2$ states by the transverse RF magnetic field (parameter $g_{3\perp}$). Far from LACs, the ratio of the $\nu_3$ and $\nu_1$ ODMR line intensities for linearly polarized RF field is given by $|M_{\text{$-3/2$, $1/2$}}|^2/|M_{\text{$-3/2$,$-1/2$}}|^2 \approx (g_{3\perp}+g_{3\parallel}/2)^2 / 4$. Using $g_{3\perp}+g_{3\parallel}/2 = 0.5$  from the fit of $\nu_{3}^{AC}$ in Fig.~\ref{fig1}(c), we obtain for the relative intensity $0.06$. It is somewhat smaller than the measured value of $\mathrm{\Delta PL_3 / \Delta PL_1} \approx 0.1$. Detailed comparison of the experimental and theoretical ODMR contrasts requires the study of the linewidths and resonance shapes in the vicinity of GSLACs, which is beyond the scope of the present paper. Given this uncertainty, we find the agreement between the theory and experiment satisfactory.

\section{All-optical magnetometry}

Having established the fine structure, we propose to use its unique properties for all-optical magnetometry. The experimental procedure is straightforward and requires no RF field. First, we tune our system in the GSLAC-2, characterized by the narrowest resonance in Fig.~\ref{fig2}(b). We then monitor the PL intensity through the lock-in in-phase photovoltage $U_X$, which is simply proportional to the deviation of the measured magnetic field from the bias field $B_{G2}$ (provided this deviation is small) [Fig.~\ref{fig4}(a)]. By applying sub-$\mathrm{\mu T}$ magnetic fields, we calibrate the lock-in signal $U_X / (B_z - B_{G2})  = 39 \, \mathrm{\mu V / \mu T}$ [Fig.~\ref{fig4}(b)]. The quadrature component $U_Y$ of the lock-in signal, being independent of the magnetic field, is used to measure the noise level. Each data point in Fig.~\ref{fig4}(b) corresponds to an integration time of $4 \, \mathrm{s}$, and the $dc$ magnetic field sensitivity is obtained to be $\delta B = 87 \, \mathrm{nT / \sqrt{Hz}}$. Indeed, magnetic fields below $100 \, \mathrm{nT}$ can be clearly resolved in Fig.~\ref{fig4}(b). 

\begin{figure}[t]
\includegraphics[width=.45 \textwidth]{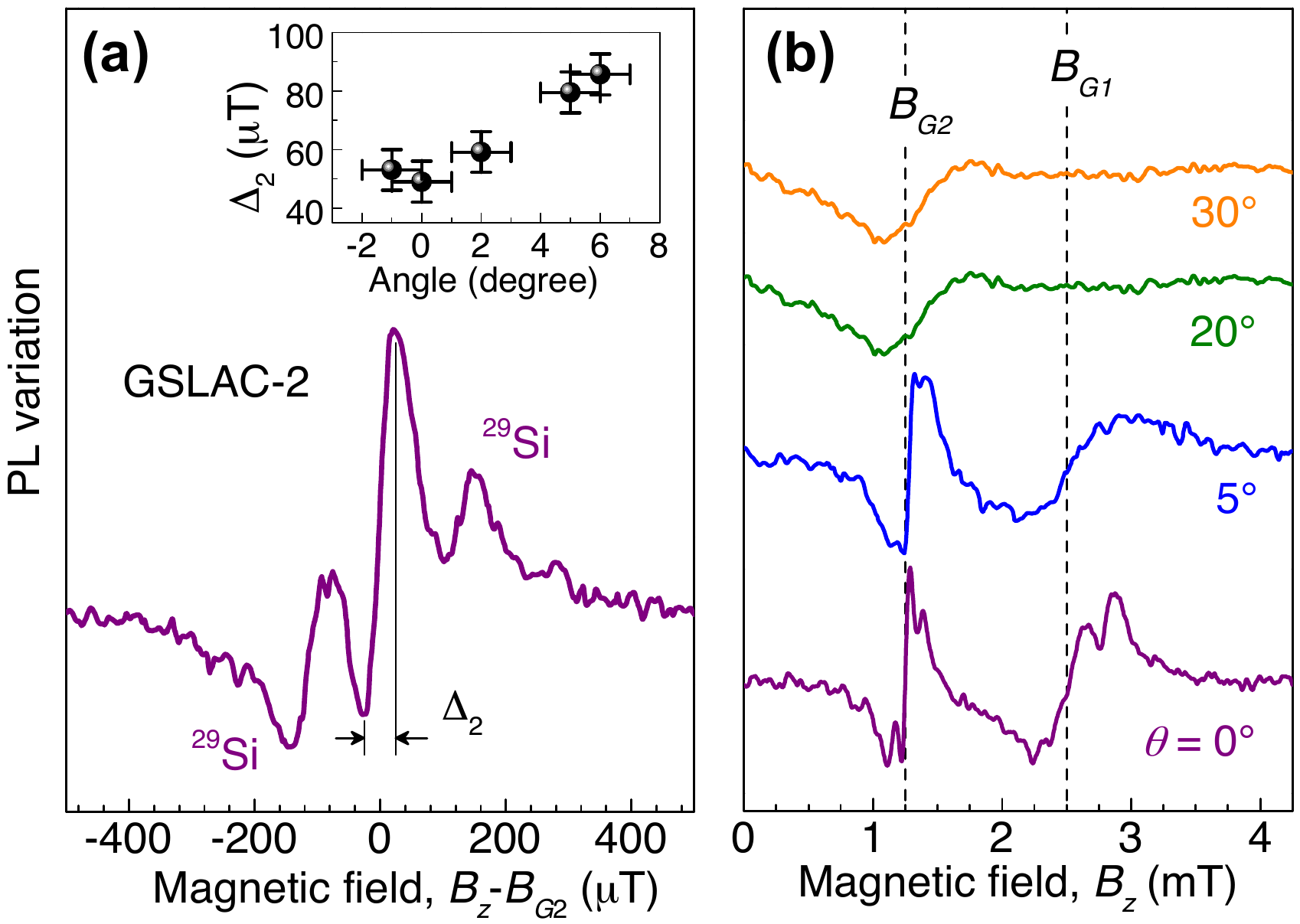}
\caption{ (a) Lock-in detection of the PL variation in the vicinity of GSLAC-2 ($B_{G2} = 1.25 \, \mathrm{mT} $), performed in 4H-SiC with natural isotope abundance.  Satellite resonances are due to the hyperfine interaction with $^{29}$Si nuclei.  Inset: variation of the peak-to-peak width as a function of the magnetic field orientation.   (b) PL variations at GSLAC-1 (at $B_{G1}$) and GSLAC-2 (at $B_{G2}$) for different magnetic field orientations with respect to the $c$-axis (polar angle $\theta$). $T = 300 \, \mathrm{K}$. The non-zero peak-to-peak width for $\theta = 0^{\circ}$ is ascribed to magnetic fluctuations of the environment (nuclear spins and paramagnetic impurities) and magnetic field alignment uncertainty (ca. $1^{\circ}$).}   \label{fig5}
\end{figure}

We use an isotopically enriched crystal to exclude possible contributions related to the hyperfine interaction with $^{29}$Si nuclei to spin Hamiltonian of Eq.~(\ref{H}). Figure~\ref{fig5} demonstrates that the proposed all-optical magnetometry can also be performed using SiC with natural isotope abundance. By alignment of the bias magnetic field along the symmetry axis with an accuracy better that one degree (the inset of Fig.~\ref{fig5}(a)), it is possible to clearly separate the spin-carrying isotope contributions. The GSLAC-2 is generally narrower and less sensitive to the magnetic field misalignment in comparison to the GSLAC-1, as can be seen in  Fig.~\ref{fig5}(b). It is indeed expected  that $\Delta_{2} < \Delta_{1}$ because the peak-to-peak width $\Delta$, as determined in \ref{fig5}(a), scales with the LAC splitting $\Lambda$ and according to Eq.~(\ref{SplittingLAC}) $\Lambda_{2} < \Lambda_{1}$. 

The dynamic range of the proposed magnetometry is relatively small, several tens of  $\mathrm{\mu T}$. It can be extended by applying a transverse magnetic field at the expense of sensitivity. On the other hand, there are many applications where large dynamic range is not required \cite{Schirhagl:2014io}. We would like to clarify that the proposed magnetometry is highly sensitive to one particular orientation of the magnetic field ($B_z$) and, therefore, designed for applications where weak magnetic variations in a certain direction must be measured with high accuracy. In order to align the magnetometer, it is necessary to conduct several preliminary measurements of magnetic field sweeps around $B_{G2}$ in differently oriented bias magnetic fields, until the maximal slope is obtained. An advantage is that the LAC can be observed even for short spin lifetimes which occur,  e.g., in excited states. The detection of the ODMR signal in those conditions may be difficult because it would require the application of highly intense RF fields. Contrary, a variation of the PL intensity at ESLAC  when the ODMR signal is not detectable has been clearly demonstrated using  GaAs/AlGaAs superlattices \cite{LAC2001}.  Our approach is robust and can be applied in a broad temperature range up to $520 \, \mathrm{K}$ [Fig.~\ref{fig4}(a)]. A crucial factor for field sensitivity is the PL intensity and stability of the pump laser. The latter factor can be compensated using a balanced detection scheme. By increasing the irradiation fluence, the $\mathrm{V_{Si}}$ density can be increased by more than two orders of magnitude  \cite{Fuchs:2015ii}, and the projected sensitivity  in this case is a few $\mathrm{nT / \sqrt{Hz}}$ within the same volume of $330 \, \mathrm{\mu m^{3}}$. Alternatively, one can use light-trapping waveguides in bigger samples \cite{Clevenson:2015cv}. For a waveguide of $3 \, \mathrm{mm} \times 3 \, \mathrm{mm} \times 300 \, \mathrm{\mu m}$ with improved collection efficiency by three orders of magnitude \cite{Clevenson:2015cv} and a $\mathrm{V_{Si}}$ density of $4 \times 10^{16} \, \mathrm{cm^{-3}}$ \cite{Fuchs:2015ii}, we estimate the projection noise limit to be below $100 \, \mathrm{fT / \sqrt{Hz}}$. In order to realize such an extremely high sensitivity, drift-compensation schemes  \cite{Wolf:2014us, Clevenson:2015cv} and magnetic noise screening  similar to that usually used for optical magnetometry based on vapour cells \cite{Shah:2007kb} are necessary to be applied.  The use of completely spin-free samples of high crystalline quality, containing $^{28}$Si and  $^{12}$C isotopes only, can lead to further improvement due to the suppression of magnetic fluctuations caused by nuclear spins.  In addition, rhombic SiC polytypes (15R) with even lower magnetic fields corresponding to the GSLACs \cite{Soltamov:2015prl} may have advantage compared to hexagonal 4H-SiC. 
 
In conclusion, we reconstruct for the first time the fine structure of the
$\mathrm{V_{Si}}$ center in 4H-SiC, quantifying its contributions. 
%It is unambiguously shown that even for the magnetic field oriented parallel to the symmetry axis four ODMR lines are expected to appear, as verified in our measurements. Furthermore, 
The presence and behaviour of two GSLACs can now be theoretically described and predicted. Our results on the spin Hamiltonian as well as the approach to study the fine structures of localized states are general. They can be directly applied for other spin-3/2 systems with the trigonal pyramidal symmetry known in solid states, such as other color centers in zinc-blende-type crystals or $\Gamma_8$-band holes in quantum dots \cite{Durnev:2013gi}. They can also be straightforwardly generalized to high-spin states such as transition metal impurities in semiconductor structures of low spatial symmetry.

These findings are directly translated to a working application, namely an all-optical magnetometry with sub-100-nT sensitivity. This is a general concept of all-optical sensing without RF fields as it can be used to measure various physical quantities, such as temperature and axial stress, through their effect on the zero-field splitting and hence on the magnetic fields corresponding to the LACs.   An intriguing possibility is to image the PL from a SiC wafer onto a CCD camera to visualize magnetic fields with temporal and spatial resolution. Our results may potentially be applied for biomedical imaging and geophysical surveying, especially when RF fields cannot be applied.

%\section*{Acknowledgments}
\begin{acknowledgments}
This work has been supported by the German Research Foundation (DFG) under grant AS 310/4, by the BMBF (Nr.~6.5~BNBest-BMBF 98) under the ERA.Net RUS Plus project "DIABASE", by RFBR Nr.~14-02-91344, RSF Nr.~14-12-00859, the RMES under agreement RFMEFI60414X0083; AVP and SAT are grateful for the support RFBR Nr. 14-02-00168, RF president grant SP-2912.2016.5, and "Dynasty" Foundation.
\end{acknowledgments}

%-------- BEGIN APPENDIX--------------

\appendix

\section{Spin transitions in the uniaxial model}\label{Unimodel}

In the uniaxial approximation, the effective Hamiltonian of a spin center has the form
\begin{equation}
H = D \left(S_z^2 - \frac54 \right) + g_{\parallel} \mu_B S_z B_z + g_{\perp} \mu_B \mathbf{S}_{\perp} \cdot \mathbf{B}_{\perp} \,,
\end{equation}
where $g_{\parallel}$ and $g_{\perp}$ are the longitudinal and transverse $g$-factors, $B_z$ and $\mathbf{B}_{\perp}=(B_x, B_y)$ are the longitudinal and transverse components of the magnetic field, respectively, and $\mathbf{S}$ is the vector composed of the spin-3/2 operators $S_x$, $S_y$, and $S_z$.

The transverse component of the magnetic field caused by inaccurate orientation of the external magnetic field along the $c$-axis
and/or originated from hyperfine interaction with nuclei couples the $+1/2$ and $-1/2$ spin states as well as the $-1/2$ and $-3/2$ spin states thereby allowing the $\nu_3$ ODMR line. Straightforward perturbation-theory calculations show that the ratio between the intensities of the $\nu_3$ and $\nu_1$ ODMR lines determined by the corresponding matrix elements of the transitions has the form
\begin{align}\label{ratio}
  & \frac{|M_{\text{$-3/2$, $+1/2$}}|^2}{|M_{\text{$-3/2$, $-1/2$}}|^2} = \left( \frac{2D}{2D-g_\parallel\mu_B B_z} \, 
	\frac{g_{\perp} B_\perp}{g_{\parallel} B_z} \right)^2 \,.
\end{align}
The experimental value of the relative OMDR contrast is about $0.1$. To obtain such a contract, e.g., for $g_\parallel\mu_B B_z = D$, one has to assume that $B_{\perp}/B_z \approx 0.16$ which corresponds to an angle of approximately $10^{\circ}$ between the total magnetic field acting upon spin centres and the $c$-axis or $B_{\perp} \approx 0.2$\,mT for $B_z = 1.25$\,mT.
Such a transverse magnetic field is an order of magnitude larger than the average nuclear field seen by the $\mathrm{V_{Si}}$ centers~\cite{Soltamov:2015prl}. The precision of the external magnetic orientation in our experiments is also by an order of magnitude better [the inset of Fig.~\ref{fig5}(a)]. For such a precision, the contrast ratio between the $\nu_3$  and $\nu_1$  ODMR lines is estimated from Eq.~\eqref{ratio} to be at most $0.001$, which is by two orders of magnitude lower than the experimentally determined ratio. 

\section{Effective Hamiltonian for the spin-$3/2$ center of the $C_{3v}$ point group}\label{Hamiltonian_construction}

We construct the effective spin Hamiltonian using the theory of group representations~\cite{BP_book}. In the $C_{3v}$ point group, there
are three irreducible representations commonly denoted as $A_1$, $A_2$, and $E$. Accordingly, all physical quantities can be 
decomposed into the irreducible representations in accordance with their symmetry properties. The magnetic field components $B_{\alpha}$ ($\alpha = x,y,z$) and all possible linearly independent combinations of the spin operator components $S_{\alpha}$ are decomposed into the
irreducible representations as follows
\begin{align}\label{reps}
A_1 :\; &  S_z^2 ; \,  S_x (3S_y^2-S_x^2) ;  \\  
A_2 :\; &  B_z ; \, S_z ; \, S_z^3 ; \, S_y(3S_x^2-S_y) ; \nonumber\\
E :\; & (B_x, B_y) ;\, (S_x,S_y) ;\, (S_yS_z,-S_xS_z) ;  (S_xS_z^2,S_yS_z^2) ;\, \nonumber \\
& (S_x^2-S_y^2,-2\{S_x,S_y\}) ;\, (2\{S_x,S_y\}S_z, (S_x^2-S_y^2)S_z) . \nonumber 
\end{align}
All other combinations of the spin operator components can be expressed via the above ones taking into account
the identity $S_x^2+S_y^2+S_z^2 = 15/4$ and the fact that the forth and higher powers of the spin-$3/2$ operator components can be reduced to the operators of lower powers. 

The effective Zeeman Hamiltonian is constructed as the sum of all possible products of the
magnetic field components and the spin operator combinations which are invariant with respect to (i) symmetry
operations of the point group and (ii) time reversal. Each invariant contribution is multiplied by a prefactor
which has a physical sense of an effective $g$-factor component. The condition (i) implies that the invariant products 
are constructed from quantities belonging to the same irreducible representation. The condition (ii) implies that
the Zeeman Hamiltonian is odd in $\mathbf{S}$. Taking both conditions into account we obtain 6 linearly independent 
contributions to the Zeeman Hamiltonian which are given in Eq.~\eqref{H}.

\section{Fine structure of spin-$3/2$ centers of the $C_{3v}$ point group}\label{Matrix_form}

The spin Hamiltonian given by Eqs.~\eqref{sum}-\eqref{H} can be rewritten in the equivalent  matrix form 
\begin{widetext}
\begin{equation}\label{H_matrix}
\mathcal{H} =  \left(
\begin{array}{cccc}
 D +\frac{3}{2} (1+\frac{g_{2\parallel}}{g_{\parallel}})  g_{\parallel} \mu_B B_z & \frac{\sqrt{3}}{2}  (1+ \frac{g_{2\perp}}{g_{\perp}})  g_{\perp} \mu_B B_- & -{\rm i}\frac{ \sqrt{3}}{2} g_{3\perp} \mu_B B_+  & -{\rm i} \frac{3}{2} g_{3\parallel} \mu_B B_z \vphantom{\dfrac12} \\
\frac{\sqrt{3}}{2}  (1+ \frac{g_{2\perp}}{g_{\perp}})  g_{\perp} \mu_B  B_+ & - D +\frac{1}{2} (1-\frac{g_{2\parallel}}{g_{\parallel}})  g_{\parallel}  \mu_B B_z & (1-\frac{g_{2\perp}}{g_{\perp}}) g_{\perp} \mu_B B_- & {\rm i} \frac{\sqrt{3}}{2} g_{3\perp} \mu_B B_+  \vphantom{\dfrac12} \\
 {\rm i} \frac{\sqrt{3}}{2} g_{3\perp} \mu_B B_-  & (1-\frac{g_{2\perp}}{g_{\perp}}) g_{\perp} \mu_B B_+ &  - D -\frac{1}{2} (1-\frac{g_{2\parallel}}{g_{\parallel}})  g_{\parallel}  \mu_B B_z & \frac{\sqrt{3}}{2}  (1+ \frac{g_{2\perp}}{g_{\perp}})  g_{\perp} \mu_B B_-  \vphantom{\dfrac12} \\
 {\rm i} \frac{3}{2} g_{3\parallel} \mu_B B_z  & - {\rm i} \frac{\sqrt{3}}{2} g_{3\perp} \mu_B B_-  & \frac{\sqrt{3}}{2}  (1+ \frac{g_{2\perp}}{g_{\perp}})  g_{\perp} \mu_B  B_+ & D -\frac{3}{2} (1+\frac{g_{2\parallel}}{g_{\parallel}})  g_{\parallel} \mu_B B_z  \vphantom{\dfrac12} 
\end{array}
\right) ,
\end{equation}
\end{widetext}
where $B_{\pm} = {B}_x \pm {\rm i} {B}_y = B_{\perp} e^{\pm {\rm i} \phi}$, with $\phi$ denoting the azimuthal angle of $\mathbf{ B}$. To derive this matrix, we use the explicit form of the spin-$3/2$ matrices
%Using the explicit form of the spin-$3/2$ matrices, 
%
\begin{eqnarray}\label{Js_x}
S_x &=& \left(
\begin{array}{cccc} 0 & \frac{\sqrt{3}}{2} & 0 & 0 \\
\frac{\sqrt{3}}{2} & 0  & 1 & 0 \\
0 & 1 & 0 & \frac{\sqrt{3}}{2} \\
0 & 0 & \frac{\sqrt{3}}{2} & 0 
\end{array}
\right) , %\;\;
\\
%\end{eqnarray}
%\begin{eqnarray}\label{Js_y}
S_y &=& \left(
\begin{array}{cccc} 0 & -\rm i \frac{\sqrt{3}}{2} & 0 & 0 \\
\rm i \frac{\sqrt{3}}{2} & 0 & - \rm i & 0 \\
0 & \rm i & 0 & - \rm i \frac{\sqrt{3}}{2} \\
0 & 0 & \rm i \frac{\sqrt{3}}{2} & 0 
\end{array}
\right) , %\;\;
\\
%\end{eqnarray}
%\begin{eqnarray}\label{Js_z}
S_z &=& \left(
\begin{array}{cccc} \frac32 & 0 & 0 & 0 \vphantom{\frac{\sqrt{3}}{2}} \\
0 & \frac12 & 0  & 0 \vphantom{\frac{\sqrt{3}}{2}} \\
0 & 0 & -\frac12 & 0 \vphantom{\frac{\sqrt{3}}{2}} \\
0 & 0 & 0 & -\frac32 \vphantom{\frac{\sqrt{3}}{2}} 
\end{array}
\right) .
%\\
\end{eqnarray}
%
%one obtains 
%

\subsection{Zeeman splitting of spin sublevels}

Application of a magnetic field along the $c$-axis leads to the splitting of
the spin sublevels. The energies of the states with the spin projections $m_S = \pm 1/2$ and $m_S = \pm 3/2$ are given by
\begin{align}
  &E_{\pm 1/2} =  -D  \pm \frac{1}{2} g_{\parallel,1/2} \, \mu_B B_z \,, \\
  &E_{\pm 3/2} =  D  \pm \frac{3}{2} g_{\parallel,3/2} \, \mu_B B_z  \,, 
\end{align}
where the effective $g$-factors are
\begin{align}
 &g_{\parallel,1/2} = g_{\parallel}-g_{2\parallel}\,, \\
 &g_{\parallel,3/2} =  \sqrt{(g_{\parallel}+g_{2\parallel})^2+g_{3\parallel}^2} %\approx  (g_{\parallel} + g_{2\parallel} + g_{3\parallel}^2/2g_{\parallel})
 \,.
\end{align}
The spin sublevel $m_S = -3/2$ crosses the spin sublevels $m_S = -1/2$ and $m_S = +1/2$ at the magnetic fields
\begin{align}
B_{G1} &= \frac{4 D}{(3g_{\parallel,3/2}-g_{\parallel,1/2}) \mu_B} \;\; \text{and}   \\
B_{G2}  &= \frac{4 D}{(3g_{\parallel,3/2}+g_{\parallel,1/2}) \mu_B} \,,
\end{align}
respectively. 

As described in the main text, application of a small additional perpendicular magnetic field $B_{\perp}$ leads to level anticrossings, GSLAC-1 and GSLAC-2 [Fig.~\ref{fig1}(a)], at $B_{G1}$ and $B_{G2}$, respectively. %The corresponding level splittings are given by

In a magnetic field applied perpendicular to the $c$-axis $B_\perp = (B_x^2 + B_y^2)^{1/2}$ and $B_z = 0$, the energy spectrum is given by
\begin{widetext}
\begin{align}\label{Eperp}
 \begin{split}
  E_{\text{$3/2$}} = \pm \frac{1}{2}(g_{\perp}-g_{2\perp})\mu_B B_\perp + \frac{1}{2}\sqrt{[2D \mp (g_{\perp}-g_{2\perp}) \mu_B B_\perp]^2 +3[(g_{\perp}+g_{2\perp})^2+g_{3\perp}^2] \mu_B^2 B_\perp^2  } \,, \\
  E_{\text{$1/2$}} = \pm \frac{1}{2}(g_{\perp}-g_{2\perp}) \mu_B B_\perp - \frac{1}{2}\sqrt{[2D \mp (g_{\perp}-g_{2\perp})\mu_B B_\perp]^2 +3[(g_{\perp}+g_{2\perp})^2+g_{3\perp}^2]\mu_B^2 B_\perp^2  } \,. 
 \end{split}
\end{align}
\end{widetext}
Particularly, for small magnetic fields ($\mu_B B_\perp \ll 2D$), the linear-in-$B_\perp$ splitting is described by the effective $g$-factors 
\begin{align}
g_{\perp,3/2} &= 0 \,, \\
g_{\perp,1/2} &= 2 ( g_{\perp}-g_{2\perp}) \,.
\end{align}

\subsection{Relation between $g$-factors in the $T_d$ point group}

The spatial arrangement of carbon atoms around the single silicon vacancy is close to the tetragonal structure, which is described by the
$T_d$ point group symmetry. The $T_d$ group symmetry is higher than the real $C_{3v}$ group symmetry of the vacancy but properly takes into account the three-fold rotation $c$-axis and allows for non-zero values of both $g_{3 \parallel}$ and $g_{3 \perp}$. Thus, one can expect that the relation between $g_{3 \parallel}$ and $g_{3 \perp}$ of the Si vacancy is close to that for the defect of the $T_d$ point group.

In the $T_d$ point group, the effective Zeeman Hamiltonian of spin-3/2 defect in the cubic axes $x' \parallel [100]$, $y' \parallel [010]$, and $z' \parallel [001]$ is described by two linearly independent parameters $g$ and $q$ and reads~\cite{IP_book}
\begin{align}\label{HTd2}
& \mathcal H_{T_d} = g\mu_B  \mathbf{ S} \cdot  \mathbf{ B}  \nonumber \\ 
& + q\mu_B \left( J_{x'}^3 B_{x'} + J_{y'}^3 B_{y'} + J_{z'}^3 B_{z'} -\frac{41}{20} \, \mathbf{ S} \cdot  \mathbf{ B} \right) .
\end{align}
In order to obtain the Hamiltonian in the axes $x \parallel [1\bar 10]$, $y \parallel [11\bar 2]$, and $z \parallel [111]$, relevant to the vacancy orientation, we use the relation between the components of the vector $\mathbf{ B}$ in two coordinate frames,
\begin{align}\label{trans}
&B_{x'} = \frac1{\sqrt2} B_x + \frac1{\sqrt6} B_y +\frac1{\sqrt3} B_z \,,  \nonumber \\
&B_{y'} = -\frac1{\sqrt2} B_x + \frac1{\sqrt6} B_y +\frac1{\sqrt3} B_z \,,  \nonumber \\
&B_{z'} = -\frac2{\sqrt6} B_y +\frac1{\sqrt3} B_z \,,
\end{align}
and similar equations for the components of the spin operator $\mathbf{ S}$. This yields
\begin{widetext}
\begin{equation}\label{HTd_matrix}
\mathcal{H}_{T_d} = \mu_B \left(
\begin{array}{cccc}
 \left(\frac{3}{2} g - \frac15 q \right)  B_z & \frac{\sqrt{3}}{2} \left( g + \frac15 q \right) {B}_- & -\sqrt{\frac38}{\rm i} q  {B}_+  & -\frac{1}{\sqrt2}  {\rm i} q  B_z \\
 \frac{\sqrt{3}}{2} \left( g + \frac15 q \right)  {B}_+ & \left( \frac{1}{2}g  + \frac35 q\right) B_z & \left(g-\frac3{10}q\right) {B}_- & \sqrt{\frac38}{\rm i} q  {B}_+    \\
 \sqrt{\frac38}{\rm i} q  B_- & \left(g-\frac3{10}q\right) {B}_+ & -\left( \frac{1}{2}g  + \frac35 q\right) B_z  & \frac{\sqrt{3}}{2} \left( g + \frac15 q \right)  {B}_-  \\
 \frac{1}{\sqrt2}  {\rm i} q  B_z  & -\sqrt{\frac38}{\rm i} q  {B}_-  & \frac{\sqrt{3}}{2} \left( g + \frac15 q \right)  {B}_+ & -\left(\frac{3}{2} g - \frac15 q \right)  B_z   \\
\end{array}
\right) \,.
\end{equation}
\end{widetext}
Comparing the Hamiltonians~\eqref{H_matrix} and~\eqref{HTd_matrix} we obtain that $g_{3\parallel}$ and $g_{3\perp}$ are related to each other by
\begin{equation}\label{g3Td}
g_{3\parallel} / g_{3\perp} = 2/3  
\end{equation}
for a defect of the $T_d$ group symmetry.

\section{Extracting the fine structure parameters}\label{Extracting_parameters}

\begin{figure}[t]
 \centering\includegraphics[width=0.44\textwidth]{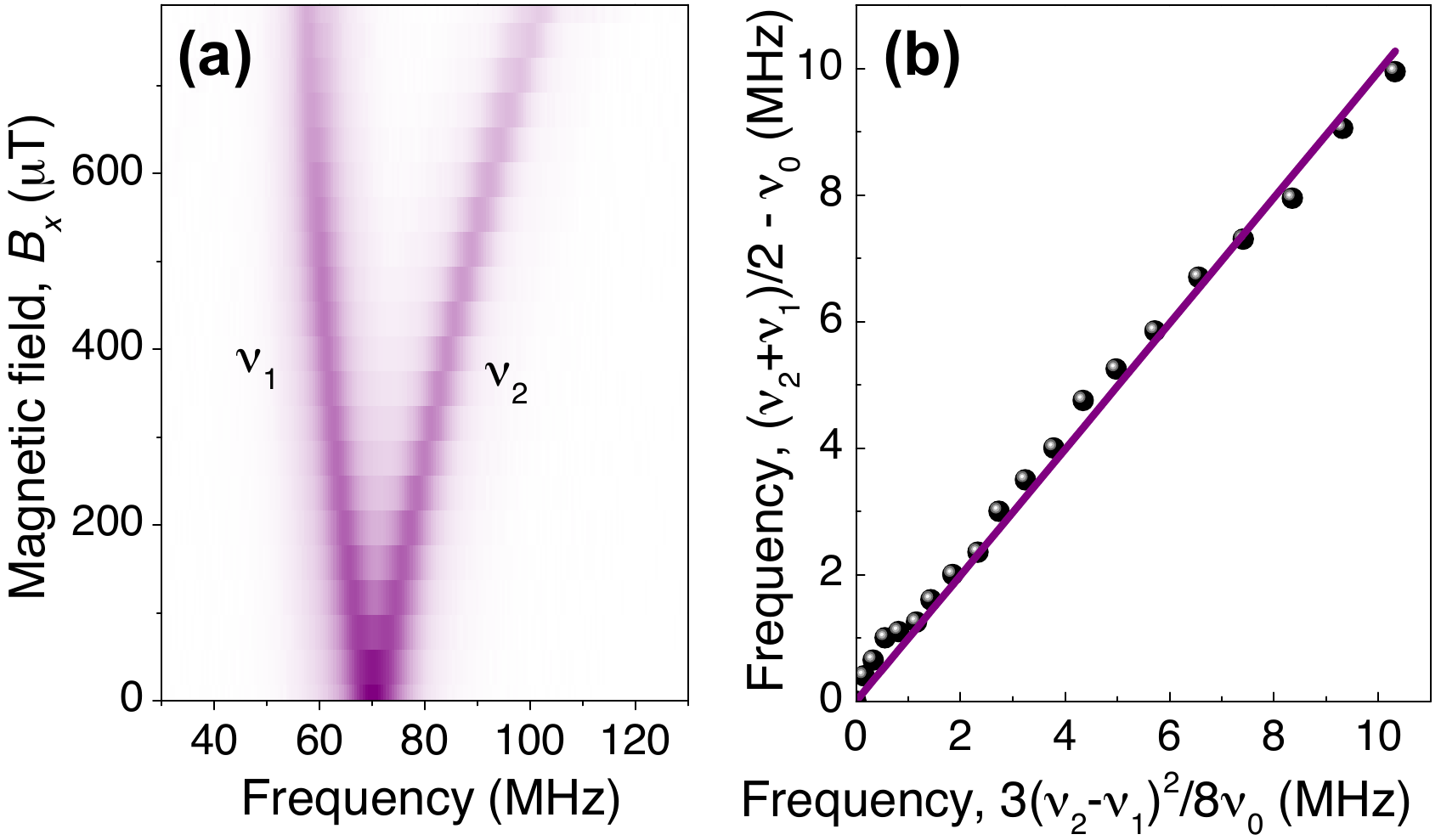}
 \caption{(a) Evolution of the ODMR spectrum as a function of the perpendicular magnetic field $B_\perp = B_x$ ($B_y = 0$). (b) Points represent the quadratic shift $(\nu_2 + \nu_1)/2 - \nu_0$ \textit{vs} the Zeeman splitting in terms of $3 (\nu_2 - \nu_1)^2/8\nu_0$. Solid line is a linear fit to Eq.~(\ref{Slope}) with a slope of $0.995 \pm 0.01$.  } 
\label{Bperp}
\end{figure}

%To obtain the value of $g_{2\parallel}$, we use the ODMR spectra in parallel magnetic field.
To obtain the value of $g_{2\parallel}$, we use the ODMR spectra, recorded in the magnetic field applied parallel to the $c$-axis. The linear shift of the ODMR lines is given by 
\begin{align}
  \nu_{1,2} & =   \nu_0  \pm \left( \frac{3}{2} g_{\parallel,3/2}  -   \frac{1}{2} g_{\parallel,1/2} \right) \mu_B B_z / h   \nonumber \\ 
 &    \approx  \nu_0  \pm(g_{\parallel} + 2 g_{2\parallel} + 3 g_{3\parallel}^2/4g_{\parallel}) \mu_B B_z / h \,, \\
  \nu_{3,4} & =   \nu_0  \pm \left( \frac{3}{2} g_{\parallel,3/2}  +   \frac{1}{2} g_{\parallel,1/2} \right) \mu_B B_z  / h  \nonumber \\ 
 & \approx \nu_0  \pm(2g_{\parallel} + g_{2\parallel} + 3 g_{3\parallel}^2/4g_{\parallel}) \mu_B B_z / h\,, 
\end{align}
where $ \nu_0 = 2D / h$. The experimentally measured ratio of the magnetic fields corresponding to the  GSLAC-2 and GSLAC-1 points, $B_{G2} / B_{G1} = 0.503 \pm 0.005$ independent of the magnetic field calibration, allows us to extract the value of $g_{2\parallel}/g_{\parallel}$  using the formula
\begin{equation}
 \frac{B_{G2}}{B_{G1}}  \approx \frac{1}{2} + \frac{3g_{2\parallel}}{4g_{\parallel}} + \frac{3g_{3\parallel}^2}{16g_{\parallel}^2} \,.  
\label{Bration}
\end{equation}
For the first iteration we neglect the term $\propto g_{3\parallel}^2$. 

To determine $g_{2\perp}$, we exploit the evolution of the ODMR spectrum in the magnetic field $B_\perp$ applied perpendicular to the $c$-axis, presented in Fig.~\ref{Bperp}(a).
From Eq.~\eqref{Eperp} we obtain the positions of the $\nu_1$ and $\nu_2$ ODMR lines up to $B_\perp^2$
\begin{align}
%  E_{\text{$3/2$}} &= D + \frac{3[(g_{\perp}+g_{2\perp})^2+g_{3\perp}^2]}{8D}  B_\perp^2 \,, \\
%   E_{\text{$1/2$}} &= -D \pm (g_{\perp}-g_{2\perp}) B_\perp - \frac{3[(g_{\perp}+g_{2\perp})^2+g_{3\perp}^2]}{8D}  B_\perp^2 \,,
  \nu_{1,2} = & \; \nu_0 \mp (g_{\perp}-g_{2\perp})\frac{\mu_B B_\perp}{h} + \nonumber \\  & \frac{3[(g_{\perp}+g_{2\perp})^2+g_{3\perp}^2]}{2\nu_0} \left(\frac{\mu_B B_\perp}{h} \right)^2 \,.
\end{align}
%From the experimental positions of the $\nu_1$ and $\nu_2$ ODMR lines (left panel in Fig.~\ref{Bperp}), 
One can represent the quadratic shift $(\nu_2 + \nu_1)/2$ \textit{vs} the Zeeman splitting $\nu_2 - \nu_1 =  2 (g_{\perp}-g_{2\perp}) B_\perp / h$, using the theoretical expression 
\begin{align}
 \frac{\nu_2 + \nu_1}{2}  - \nu_0 &= \frac{3[(g_{\perp}+g_{2\perp})^2+g_{3\perp}^2]}{8 (g_{\perp}-g_{2\perp})^2} \,\frac{(\nu_2 - \nu_1)^2}{\nu_0}   \nonumber \\ 
 & \approx \left(  1 +  \frac{4g_{2\perp}}{g_{\perp}} + \frac{g_{3\perp}^2}{g_{\perp}^2} \right) \frac{3 (\nu_2 - \nu_1)^2}{8  \nu_0}  
\label{Slope}
\end{align}
From the linear fit of the experimental data in Fig.~\ref{Bperp}(b), we obtain the value of $g_{2\perp}/g_{\perp}$.
Again, for the first iteration we neglect the term $\propto g_{3\perp}^2$.   

Finally, we use the GSLAC-1 and GSLAC-2 splittings, described by Eqs.~\eqref{SplittingLAC}, to determine the values of $g_{3\parallel}$ and $g_{3\perp}$. By iterating the described procedure, we obtain the values of the $g$-factors, as presented in table~\ref{AllParam}.

\section{Electron spin resonance}

The spin transition rates induced by the RF magnetic field $\mathbf{ B}_1$ are determined by the matrix elements of Hamiltonian~\eqref{H_matrix}. For a static magnetic field $ \mathbf{B}\parallel z$, the matrix elements of the transitions up to linear in $g_{2\parallel}$, $g_{2\perp}$, $g_{3\parallel}$, and $g_{3\perp}$ terms are given by Eqs.~\eqref{M1}-\eqref{M2} in the main text, and the full expressions have the form
\begin{widetext}
\begin{align}\label{M_epr_12}
  & M_{\text{$-3/2$, $-1/2$}} = \sqrt\frac38 \left[(g_{\perp}+g_{2\perp}) \sqrt{1+\frac{g_{\parallel}+g_{2\parallel}}{g_{\parallel,3/2}}} + \frac{g_{3\parallel}g_{3\perp}}{\sqrt{g_{\parallel,3/2}}\sqrt{g_{\parallel}+g_{2\parallel}+g_{\parallel,3/2}}}\right]\mu_B (B_{1,x} + {\rm i} B_{1,y}) \,, \\
    & M_{\text{$3/2$, $1/2$}} = \sqrt\frac38 \left[(g_{\perp}+g_{2\perp}) \sqrt{1+\frac{g_{\parallel}+g_{2\parallel}}{g_{\parallel,3/2}}} + \frac{g_{3\parallel}g_{3\perp}}{\sqrt{g_{\parallel,3/2}}\sqrt{g_{\parallel}+g_{2\parallel}+g_{\parallel,3/2}}}\right]\mu_B (B_{1,x} - {\rm i} B_{1,y}) \,,\\
 %\end{align}
 %\begin{align}\label{M_epr_32}
 & M_{\text{$-3/2$, $1/2$}} = -{\rm i} \sqrt\frac38 \left[ \frac{g_{3\parallel}(g_{\perp}+g_{2\perp})}{\sqrt{g_{\parallel,3/2}}\sqrt{g_{\parallel}+g_{2\parallel}+g_{\parallel,3/2}}} + g_{3\perp}\sqrt{1+\frac{g_{\parallel}+g_{2\parallel}}{g_{\parallel,3/2}}}\right]\mu_B (B_{1,x} - {\rm i} B_{1,y}) \,,\\
  & M_{\text{$3/2$, $-1/2$}} = -{\rm i} \sqrt\frac38 \left[ \frac{g_{3\parallel}(g_{\perp}+g_{2\perp})}{\sqrt{g_{\parallel,3/2}}\sqrt{g_{\parallel}+g_{2\parallel}+g_{\parallel,3/2}}} + g_{3\perp}\sqrt{1+\frac{g_{\parallel}+g_{2\parallel}}{g_{\parallel,3/2}}}\right]\mu_B (B_{1,x} + {\rm i} B_{1,y})  \,.
\end{align}
 The matrix element for $(-1/2 \leftrightarrow +1/2)$ has the form 
%\begin{equation}
$M_{\text{$1/2$, $-1/2$}} = (g_\perp - g_{2\perp}) \mu_B (B_{1,x} - {\rm i} B_{1,y})$, 
%\end{equation}
while the spin transition $(-3/2 \leftrightarrow  +3/2)$ is forbidden for $ \mathbf{B}\parallel z$.
\end{widetext}

%\end{widetext}

%-------- END APPENDIX--------------

%\bibliography{SiC-Anticrossing}

%

\end{document}